\documentstyle[12pt,epsfig]{article}
\begin{document}
\title{
$e^+ e^- \to e^- {\bar \nu_e} u {\bar d}$ \\
from LEP to linear collider energies}
\author{ Y. Kurihara$^a$, D. Perret-Gallix$^b$, and Y. Shimizu$^a$\\ \\
$^a$National Laboratory for High Energy Physics,\\
Oho 1-1, Tsukuba, Ibaraki 305, Japan\\
$^b$Laboratoire d'Annecy-le-Vieux de Physique des Particules \\
(LAPP), B.P. 110, F-74941 Annecy-le-Vieux Cedex, France}

\date{}

\maketitle

\begin{abstract}
The complete tree level cross-section for the process
$e^+e^- \to e^- \bar\nu_e u \bar{d}$ is computed using
the GRACE system, a program package for automatic
amplitude calculation.
Special attention is brought to the gauge violation
problem induced by the finite width of the $W$-boson.
The {\it preserved gauge scheme} is introduced and
an event generator including double-resonant,
single-resonant and non-resonant diagrams with no need for a
cut on the electron polar angle is built.
Below threshold, the single $W$ and non-resonant diagrams
give a substantial contribution to the total cross-section,
at linear collider energies, the cross-section, for small electron
polar angles, is simply dominated by these contributions.
Since no cut needs to be applied to the electron, the generator
can be used to estimate background for searches involving jets and missing
energy. A monojet event rate estimation based on
this process at LEP-I energy is discussed.
\end{abstract}
\thanks{~ \\
{}~~~~~~~~~~~~~~~ To be submitted to Physics Letters B.}

\maketitle
\newpage

\section{Introduction}
Precise event generators for processes involving $W$, $Z$, $\gamma$
and leading to four fermions in the final state
are necessary for a good understanding of the three-boson
coupling at LEP and at future linear collider C.M. energies.
Four-fermion final state can be produced
by double (heavy boson) resonant
diagrams, single resonant diagrams or non-resonant diagrams.
Single $W$ processes include
$e^+e^- \to l^- \bar\nu_l W^+$ where $l=e,\mu$.
The $l=e$ case differs from the
$l=\mu$ one by the existence of the diagrams where $\gamma$,
$Z^0$ or even $W$ are exchanged in the $t$-channel.
Most of these diagrams have a $\gamma$-$\gamma$ like behavior such as
a strong forward electron peak, the difference lying in the
propagator mass and width.
These diagrams will be called the $t$-channel diagrams
hereafter in this paper.
Since the $W$-boson is unstable and decays,
final state like
$e^+e^- \to e^- \bar\nu_e u \bar{d}$ will be actually observed.
The diagrams involved in $e^+e^- \to e^- \bar\nu_e u \bar{d}$ can be
grouped into the $s$-channel (Fig.1)
and $t$-channel classes (Fig.2), each group
forms a gauge invariant set. In regard to the
$W$-resonance, they can also be characterized as
double-resonant, single-resonant, and non-resonant diagrams.
%As far as the $W$-mass reconstruction is concerned, the single- and
%non-resonant diagrams should
%be counted as background processes, and they will give rise to
%uncertainties on the overall mass and width measurement.
The single and non-resonant diagrams play an important role on several
studies, including the $W$ mass determination, the anomalous
couplings limits and the background estimates in the search of new particles.

Recently event generators
for four-fermion (+ photons) processes have been developed \cite{gene}.
However, in order to cope with the complexity of the calculations,
some approximations are always introduced.
On the contrary, the GRACE package can automatically produce
the complete set of the tree level diagrams involved in the
process while taking into account fermion masses.

In a previous work \cite{sotchi},
a complete calculation of two typical processes of four-fermion final
states in $e^+e^-$ collisions, $e^+e^- \to e^-\bar\nu_e u \bar{d}$ and
$e^+e^- \to \bar{u} d u \bar{d}$, has been presented for the first time.
We have pointed out that the single and
non-resonant diagrams play an important role below the $W$-pair
threshold. Their contribution, for instance,
reaches 27\% for the $\bar{u} d u \bar{d}$ process
for $\sqrt{s}=150$ GeV and
gives a non-negligible effect even at higher energy (4.4\%
for $e^-\bar\nu_e u \bar{d}$ at $\sqrt{s}=190$ GeV).
Below the threshold only the off-shell double-resonant amplitudes
compete with the single- and non-resonant ones, so that
the relative contribution of the latter becomes quite large.
Besides a set of experimental loose constraints,
a cut on the polar angle of the final electron with respect to
the initial electron direction, $\theta_e$, was applied to avoid
a significant gauge violation appearing in the subset of $\gamma$-$W$
diagrams.

In this paper, we analyze the effect of the gauge violating term
due to the finite width of the $W$-boson.
The so-called {\it preserved gauge scheme} is introduced
to overcome this problem.
The total $e^+e^- \to e^-\bar\nu_e u \bar{d}$ cross-section
from below the $W$-pair threshold
up to linear collider energy,  with no cut on the final
electron, is presented.
As no constraint need to be applied on the electron kinematic, this
generator can be used to estimate the background to signal requesting jets
and missing energy like the search for new particles. As an example,
the contribution of this process for mono-jet production
at LEP-I is estimated in the last section of this paper.

\section{The GRACE system}
The GRACE system \cite{grace}
has been developed to perform the very lengthy computations involved in the
study of high energy reactions. The GRACE package is a complete
set of tools for computing tree level processes.
All the usual steps occurring in a given computation
are covered: from the process specification
to the event generator. It is composed
of three components: the diagram generator, the
matrix element builder using helicity amplitudes from the
CHANEL \cite{chanel} library and the multi-dimensional phase space
integration package BASES
\cite{bases} associated with the event generator
SPRING \cite{bases}.

Fermion masses are properly taken into account in the helicity amplitudes.
The boson width is introduced into the gauge propagator when
the denominator may vanish for positive squared momentum transfer.
A gauge invariance checking program is automatically built by
the system.

The results presented hereafter have been obtained using the following
set of parameters:
\begin{eqnarray*}
M_Z & = & 91.1~{\rm GeV}\\
\Gamma_Z & = & 2.534~{\rm GeV}\\
\alpha & = & 1/137\\
\sin^2\theta_W & = & 1-(M^2_W/M^2_Z)\\
M_W & = & 80 ~{\rm GeV}\\
m_u &=&m_d~=~0.1~{\rm GeV}.
\end{eqnarray*}

The $W$ width is taken from the Particle Data Group Table:
$\Gamma_W=2.25$ GeV. The gauge boson ($W,Z$)
widths are assumed to be constant in the calculation. Furthermore some
realistic experimental cuts have been introduced:
\[
\left\{
\begin{array}{ccc}
0^\circ     &< \theta_{e^-}      <& 180^\circ ~~~{\rm case-a} \\
8^\circ     &< \theta_{e^-}      <& 172^\circ ~~~{\rm case-b} \\
\end{array}
\right\}
,~
\begin{array}{ccc}
8^\circ     &< \theta_{u,\bar d} <& 172^\circ \\
\end{array}
,~
\begin{array}{ccc}
E_{u,\bar d} &> 1~{\rm GeV}
\end{array}
\]
where $\theta_{e^-}$ is the final state electron polar angle,
measured from the incident $e^-$ beam,
$\theta_{u,\bar d}$ are the similar angle for the $u$ and $\bar d$ quark
and $E_{u,\bar d}$ are
the energies of final $u$ and $\bar d$ quarks.

The gauge invariance of the amplitude without particle width is
checked numerically by
a random selection of the boson gauge parameters at several points
of the phase space. The errors are within the precision of the numerical
calculation (typically less than ${\cal O}(10^{-12})$ in double
precision).

\section{Effect of the gauge violating term}

The violation of the gauge invariance at the tree level is
due to the
introduction of the $W$-boson finite width. The first four diagrams
in Fig.2, the so-called $\gamma$-$W$ diagrams, give the dominant contribution
to the $t$-channel amplitude,
the $e^-$ being scattered in the forward
direction. Since a large cancellation occurs among
$\gamma$-$W$ diagrams \cite{singlt},
the gauge violating terms lead to a strongly divergent
cross-section at small electron polar angle.
It blows up by about six orders
of magnitude at $\theta_e \approx 0$ when the width is introduced
directly in the propagator (Fig.3, dashed line).

To see how the effect of the gauge violating term arises,
we examine the total amplitude of the $\gamma$-$W$ diagrams.
Since there is only one $W$ propagator whose
four-momentum transfer squared is positive, the amplitude without the
$W$-width can be written as:
\begin{eqnarray}
{\cal M }&=&-\frac{e}{k^2}l_\mu T_\mu, \\
 T_\mu &=&   \frac{r_\mu}{q^2-M_W^2} +n_\mu,\\
 l_\mu &=&{\bar u}(p')\gamma_\nu u(p)
[g_{\mu \nu}+(\xi -1) k_\mu k_\nu/k^2],\\
            &=&{\bar u}(p')\gamma_\mu u(p),
\end{eqnarray}
where $p_\mu$($p'_\mu$) is the four-momentum of the initial (final)
electron, $k_\mu=p_\mu-p'_\mu$, the momentum of the virtual
photon, $q_\mu$, the $W$-propagator four-momentum transfer ($q^2 > 0$) and
$r_\mu$($n_\mu$), the single-resonant (non-resonant) diagram amplitudes.
If the width, $\Gamma_W$, is introduced, one obtains:
\begin{eqnarray}
T_\mu \to T_\mu' &=& \frac{r_\mu}{q^2-M_W^2+i M_W \Gamma_W} + n_\mu \\
 &=& \frac{d_\mu + i M_W \Gamma_W n_\mu}{q^2-M_W^2+i M_W \Gamma_W}, \\
 d_\mu &=& r_\mu + (q^2 - M_W^2) n_\mu.
\end{eqnarray}
The square of the electron current, after averaged over
spin states, is:
\begin{equation}
L_{\mu \nu}=\overline{\sum_{spin}} l_\mu l_\nu^*
=2[p_\mu p_\nu' + p_\nu p_\mu'+\frac{k^2}{2} g_{\mu \nu}].
\end{equation}
Then the squared amplitude is given by:
\begin{eqnarray}
| {\cal M} |^2 &=& \frac{2 e^2}{(k^2)^2}
\frac{(L_{\mu \nu}d_\mu d_\nu^*) + M_W^2 \Gamma_W^2(L_{\mu \nu}
n_\mu n_\nu^*)}{(q^2-M_W^2)^2+M_W^2 \Gamma_W^2},
\end{eqnarray}
(note that $r_\mu$ and $n_\mu$ are real numbers at the tree level).
The first term in numerator is the gauge invariant part;
when $k^2 \to 0$, it behave as
$L_{\mu\nu}d_\mu d_\nu^* \to {\cal O}(k^2)$.
The second term is the gauge violating one as
$L_{\mu\nu}n_\mu n_\nu^* \to {\cal O}(1)$ when $k^2 \to 0$.
By integrating over $k^2$, the former gives the well-known
$\log(s/m_e^2)$ dependence of the total cross-section.
The latter, the gauge violating term,
however, dominates the total cross-section after integration, as it
does not compensate the photon propagator $\propto 1/(k^2)^2$.
It is clear that this term is responsible for the divergence of
the cross-section in the small angle region (Fig.3, dashed line).
It should be emphasized that this behavior does not depend on the
gauge parameter of the $W$ propagator.

The introduction of the $W$ width, yet preserving the gauge invariance
of the amplitude, can be achieved by applying the following method.

The current $l_\mu$ is replaced by the momentum $k_\mu$,
gauge invariance implies:
\begin{eqnarray}
k_\mu T_\mu &=& \frac{k \cdot r}{q^2-M_W^2} + k \cdot n, \\
            &=& \frac{k \cdot d}{q^2-M_W^2}=0.
\end{eqnarray}
Then $d_\mu$ is a gauge invariant quantity. $T_\mu$ can be
expressed in term of $d_\mu$ as:
\begin{equation}
T_\mu = \frac{d_\mu}{q^2-M_W^2}.
\end{equation}
If the particle width is introduced at this stage,
the amplitude can be cast into the form:
\begin{equation}
T_\mu \to T_\mu'' = \frac{d_\mu}{q^2-M_W^2+i M_W \Gamma_W}.
\end{equation}
This amplitude is apparently gauge invariant
as the divergent term discussed previously has disappeared \cite{zeppenfeld}.
The total cross-sections based on this amplitude do not diverge even in
the small angle region as shown by the solid line in Fig.3.
This scheme, the {\it preserved gauge scheme},
can be interpreted as follows; the tree-level amplitude
is gauge invariant but divergent at $q^2=M_W^2$. To avoid this divergence,
an imaginary part of a higher-order is introduced into
the propagator denominator as a particle width. However this order mixing
causes the violation of the gauge invariance at the tree-level
and gives rise to the cross-section divergence.
To preserve the gauge invariance, an additional term
is added to the amplitude on the analogy of the counterterms,
it gives the higher order correction to the leading term.

In order to check the validity of the method, one can approach this result
by applying a method which
minimize the gauge violating term. Let us remind a usual trick used
for the tensor
$L_{\mu \nu}$. Let's assume, first, that $T_\mu$ is gauge invariant.
One can replace $L_{\mu \nu}$ by:
\begin{eqnarray}
L_{\mu \nu}  \to L_{\mu \nu}' = 4 p_\mu p_\nu +  k^2  g_{\mu \nu}.
\end{eqnarray}
In this equation, the first term is responsible for the blow-up of the cross
section. Because of the gauge invariance,
one can further replace the vector $p_\mu$ by:
\begin{eqnarray}
p_\mu \to P_\mu = p_\mu-(p_0/k_0)k_\mu,
\end{eqnarray}
where $p_0$ and $k_0$ are the 0-th components of the four-momenta
$p_\mu$ and $k_\mu$, respectively.
%This replacement is legitimate again because of the gauge invariance
%Eq.(10).
By substituting $p_\mu$ in Eq.(15) and dropping $k_\mu$, one gets
\begin{eqnarray}
L_{\mu \nu}'  \to L_{\mu \nu}'' = 4 P_\mu P_\nu + k^2 g_{\mu \nu}.
\end{eqnarray}
It is known that a product $P \cdot A$ of $P$ with an arbitrary vector $A$
can be expressed by a sum of terms
proportional to either $m_e^2$, $1-\cos{\theta_e}$
or $\sin{\theta_e}$.
Hence in the region $\theta_e \approx 0$, both $P \cdot d$ and
$P \cdot n$ behave like $k^2$, because $1-\cos{\theta_e}$ vanishes
almost like $k^2$.

Then, one possible way to get rid of the large gauge violation is to use
the current $L_{\mu \nu}''$, instead of the original one
$L_{\mu \nu}$ in Eq.(9). The new amplitude squared becomes:
\begin{equation}
| {\cal M} |^2 \to \frac{2 e^2}{(k^2)^2}
\frac{(L_{\mu \nu}''d_\mu d_\nu^*) + M_W^2 \Gamma_W^2(L_{\mu \nu}''
n_\mu n_\nu^*)}{(q^2-M_W^2)^2+M_W^2 \Gamma_W^2}.
\end{equation}
The gauge violating term $L_{\mu \nu}''n_\mu n_\nu$
behaves as ${\cal O}(k^2)$ for vanishing $k^2$, and it should not
induce violent divergence.
Since the non-resonant diagrams have a $W$-boson space-like propagator,
one may assume roughly that ${\cal O}(r_\mu) \sim {\cal O}(n_\mu M_W^2)$.
Hence the gauge violating term is proportional to
$(\Gamma_W/M_W)^2\sim10^{-3}$. This is small compared to the gauge
invariant term $(L_{\mu \nu}''d_\mu d_\nu^*)$
and can be neglected.

Using this last method, the cross-section turns out
to be 8.63$\times 10^{-2}$ pb (case-a at $\sqrt{s}=180$ GeV)
as seen in Fig.3 (indicated by the arrow).
This is only 5\% larger than the cross-section obtained by the
{\it preserved gauge scheme} (8.24$\times 10^{-2}$ pb).

\section{Threshold behavior}
In the following, the computations are performed using the
{\it preserved gauge scheme}.
The contribution from the $t$-channel
diagrams strongly depends on $\theta_e$ cut
as shown in Fig.4. If we require the electron polar angle to be in the range
$172^\circ > \theta_e > 8^\circ$(140 mrad) case-b,
the double-resonant
diagrams are dominant (96\% at $\sqrt{s}=180$ GeV) but
their contributions are reduced down to 83\%
if the electron angular cut is set to a vanishing value (case-a).

The threshold behavior of $e^+e^- \to e^- \bar\nu_e u \bar{d}$
total cross-section is presented in Fig.5 together with
one high energy point ($ \sqrt{s}=500$ GeV).
%The $W$-width is introduced using the {\it preserved gauge scheme}.
With no electron angular cut, case-a, one observes
a large increase in the cross-section due to the $t$-channel diagrams,
although when such a cut is applied, case-b, the effect remains substantial
only above threshold.
The relative contribution of the
double-resonant diagrams is shown in Fig.6.
At $\sqrt{s}= 500$ GeV, the share of the double-resonant
diagrams is only 17\% without electron tagging (case-a), while it amounts to
55\% when the electron is tagged (case-b).

For the $W$ mass measurement, it should be noted that using the
threshold scanning method,
a precise electron tagging is necessary to deal with
the $t$-channel contribution. When the direct reconstruction of 2-jets
is used, the invariant mass distribution of a quark pair gets a
low mass tail from
the non-resonant diagrams. The effect of the electron angular cut is to
suppress the contribution from those diagrams,
as shown in Fig.7.

In the studies of the anomalous couplings of the gauge bosons,
the process $e^+ e^- \to e \nu_e W$ is considered.
However the process to be measured is
$e^+ e^- \to e \nu_e q {\bar q}$ \cite{anomc} which includes the
effect of the non-resonant diagrams. The contributions
of $t$-channel diagrams (Fig.2) have been computed for both processes
$e^+ e^- \to e \nu_e W$ and $e^+ e^- \to e \nu_e q {\bar q}$,
with no cut, a 14\% discrepancy is found in the total cross-section at
$\sqrt{s}=180$ GeV, which comes from the small angle region of the final
electron.
\[1- \frac{\sigma (ee \to e\nu W)*Br(W \to ud)}
{\sigma (ee \to e \nu ud)}=0.14\]
This effect should definitely be taken into account in
these studies.

The {\it preserved gauge scheme} can also be applied to
different processes,
like those appearing at electron-electron collider. For example, the process
$e^-e^- \to e^-W^-\nu_e$  \cite{barger}
involves only the $t$-channel diagrams at the tree level.

A comparison with the four-fermion Monte-Carlo {\sc EXCALIBUR} \cite{excalibur}
has been performed for the $e^+ e^- \to e^- {\bar \nu_e} u {\bar d}$ channel
as an independent test of our work.
However such a comparison make sense only when a cut is applied on the
electron polar angle as {\sc EXCALIBUR} does not treat the forward divergence.
Under this condition, perfect agreement is obtained with our generator
without the {\it preserved gauge scheme}.
Identical divergent behavior is
found when the cut is decreased down to almost zero.
The net result of our study is the extension of the domain of validity
of the generator for massive fermions and for vanishing electron polar angle.

\section{Monojet event rate at LEP}

The ALEPH collaboration at LEP has reported the observation of
three monojet events,
one leptonic and two hadronic,
from a data sample of 82 $pb^{-1}$ \cite{monojet}
recorded at and close to the $Z^\circ$ peak.
Although the event rate is consistent with the expectation from
$e^+e^- \to \gamma^* \nu {\bar \nu}$ with $\gamma^* \to f {\bar f}$,
the large observed monojet mass and transverse momentum are quite
unlikely in this process.
Even when diagrams from $Z$ or $W$
exchanges are taken into account, the probability of their occurrence is
about 5\%.

For the $W$ exchange process, $e^+e^- \to e \nu f {\bar f}'$, only
four types of diagrams are taken into account, double-resonant
and single-resonant
diagrams in $s$-channel and single-resonant diagrams in $t$-channel.
Non-resonant diagrams were not included as no generator was available
at that time. These diagrams are expected to give a large
contribution to the total cross-section at energies below the $W$-pair
threshold, moreover the mass and transverse momentum distribution
are expected to peak at higher values.

The expected number of events from
$e^+e^- \to e^{\pm(} {\bar \nu_e}^) q {\bar q}'$ and the probability
to produce the observed jet masses and transverse momenta have been estimated
by applying the ALEPH cuts at the generator level.
\begin{enumerate}
\item $\theta_e < ~25~{\rm mrad}.$
\item Polar angle of $q {\bar q}'$ system $< 25.8^\circ$.
\item Neither $q$ nor ${\bar q}'$ goes backward of
$q {\bar q}'$ system.
\item $E_{q {\bar q}'} > 1.3$ GeV.
\end{enumerate}
The total cross-section of the
$e^+e^- \to e^{\pm(} {\bar \nu_e}^) q {\bar q}'$
processes ($q {\bar q}'=u{\bar d}$ and $c{\bar s}$) at $\sqrt{s}= 91.1$ GeV
with the above cuts is $7.2 \times 10^{-3}~pb$, which corresponds
to 0.59 events for a integrated luminosity of 82 $pb^{-1}$.
%The calculation have been done including all diagrams at tree level and
%using the {\it preserved gauge scheme} to introduce the $W$ width.
The expected
event distribution of $q {\bar q}'$ invariant masses and transverse momenta
are shown in Figs.8 and 9, respectively with
the same binning as Fig.3 in ref.\cite{monojet}.
The observed jet masses and
transverse momenta of the hadronic
monojet ($M~=$ 3.2, 5.3 GeV and ${\it p_T}$ = 6.6, 18.5 GeV)
are distributed around
the peak of the distributions.
The probability defined in ref.\cite{monojet} is
calculated to be 0.95. The observed events are compatible with
the expectation obtained from
$e^+e^- \to e^{\pm (} {\bar \nu_e}^) q {\bar q}'$ processes. However
detailed calculations with hadronization and detector simulation are
needed to give a more precise estimation.

\section{Summary and Conclusions}

The complete tree level cross-section of the process
$e^+ e^- \to e^- {\bar \nu_e}u{\bar d}$ has been computed using the
GRACE package. It is shown that the naive method of introducing the
$W$ width induces an unphysical divergent total cross-section
due to the gauge violation among the $t$-channel tree-level diagrams.
This effect is sizable when the electron scattering angle is less than
a few degrees. Two {\it ad hoc} methods are discussed, one curing
definitely
the violent blow up of the amplitude (the {\it preserved gauge scheme})
and the other strongly reducing this violation.

The cross-sections as a function of the CM energies without any cut
on the final electron are presented. We conclude that the
approximation based only on a part of the diagrams (the double-resonant
diagrams and/or the single-resonant $\gamma$-$W$ diagrams) is
not sufficient to give precise predictions neither
at threshold energy nor at high energy.
For the $W$ mass determination based on the $W$ pair threshold scanning,
the complete calculation should be used and
a precise measurement of the electron should be made.
The single-resonant diagrams
can be used for the $W$ mass measurement by the 2 jet reconstruction
method, although the kinematic constraints are much weaker.
Observed jet masses and
transverse momenta of hadronic
monojet seen by the ALEPH collaboration at LEP are not in contradiction with
the expectation from
$e^+e^- \to e^{\pm(} {\bar \nu_e}^) q {\bar q}'$ processes when
naive experimental cuts are applied at parton level.

\vskip 1.0cm
\begin{center}
   {\bf Acknowledgment}
\end{center}
\vskip 0.5cm

We would like to thank our colleagues in Minami-Tateya theory group
of KEK and those in LAPP for valuable discussions and encouragement.
Particularly we are indebted to T. Kaneko, H. Tanaka and T Ishikawa
for their help in GRACE system and S. Kawabata for providing a high quality
diagram drawing package.
We would like also to thank J.~Kubo, I. Ginzburg and F. Cuypers for
fruitful discussions on the treatment of the particle width.

This work has been done in the framework of the KEK-LAPP
collaboration supported in part
by the Ministry of Education, Science and Culture(Monbusho) under
the Grant-in-Aid for International Scientific Research Program
No.04044158 in Japan and by le Centre National de la Recherche
Scientifique (CNRS/IN2P3) in France.

%\newpage
%%%%%%%%%%%%%%%%%%%%%%%%%% ref %%%%%%%%%%%%%%%%%%%%%%%%%%%%%%%%%%%%%%%

%\newpage
%%%%%%%%%%%%%%%%%%%%%%%% fig 1 %%%%%%%%%%%%%%%%%%%%%%%%%%%%%%%%%%%%%%%%
\begin{figure}[h]
\begin{center}
%\vspace*{-1cm}
\mbox{\epsfig{file=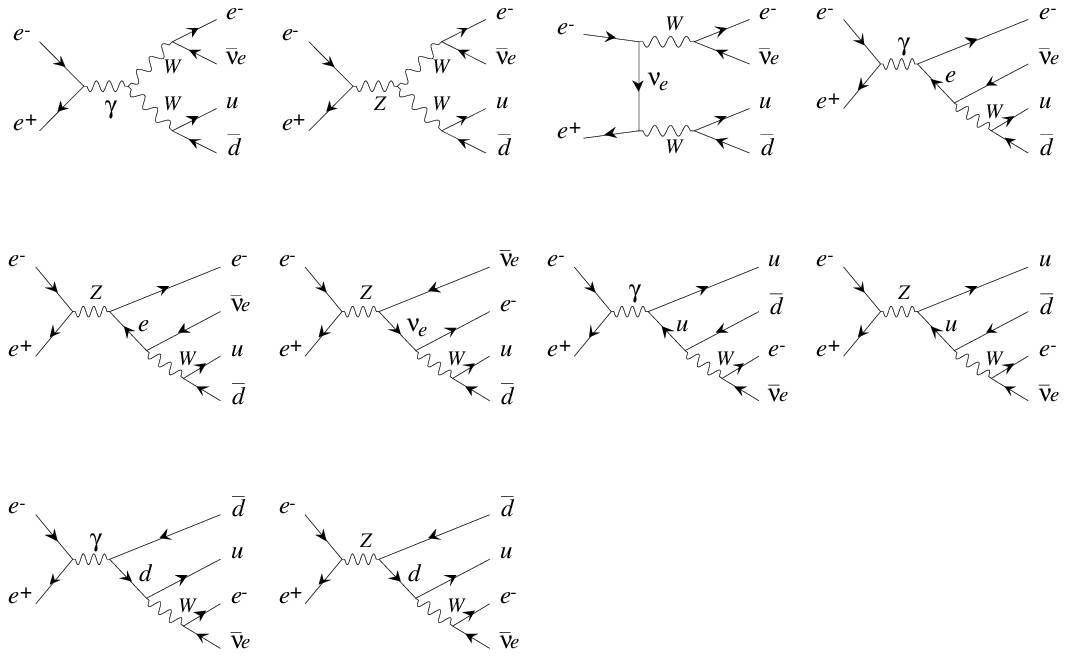}}
\caption{The $s$-channel diagrams of the
 $e^+e^- \to e^- \bar\nu_e u \bar{d}$ process in unitary gauge.
 The first three diagrams in the first row are double-resonant
diagrams.}
\end{center}
\end{figure}
%%%%%%%%%%%%%%%%%%%%%%%% fig 2 %%%%%%%%%%%%%%%%%%%%%%%%%%%%%%%%%%%%%%%%
\begin{figure}
\begin{center}
\mbox{\epsfig{file=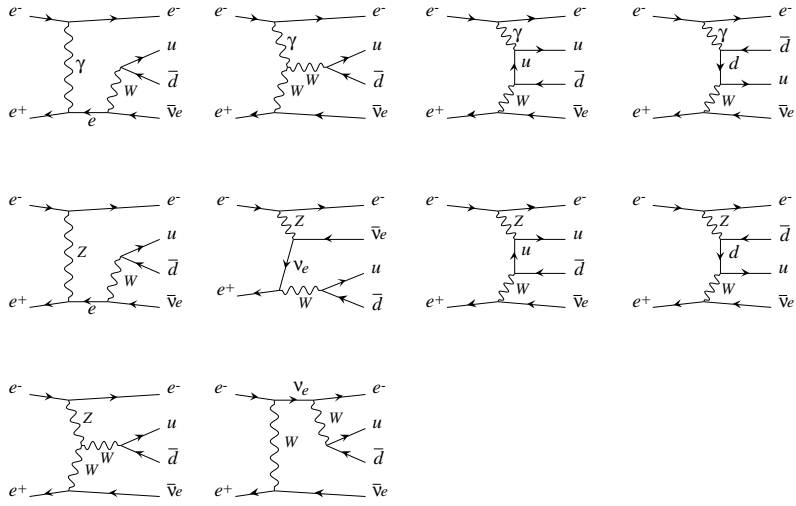}}
\caption{The $t$-channel diagrams of
 $e^+e^- \to e^- \bar\nu_e u \bar{d}$ process in unitary gauge.
 The first and second columns show
 the single-resonant diagrams and the rest shows
 the non-resonant diagrams. Diagrams in first row($\gamma$-$W$ processes)
 give the dominant contribution among $t$-channel diagrams.}
\end{center}
\end{figure}
%%%%%%%%%%%%%%%%%%%%%%%% fig 3 %%%%%%%%%%%%%%%%%%%%%%%%%%%%%%%%%%%%%%%%
\begin{figure}
\begin{center}
\mbox{\epsfig{file=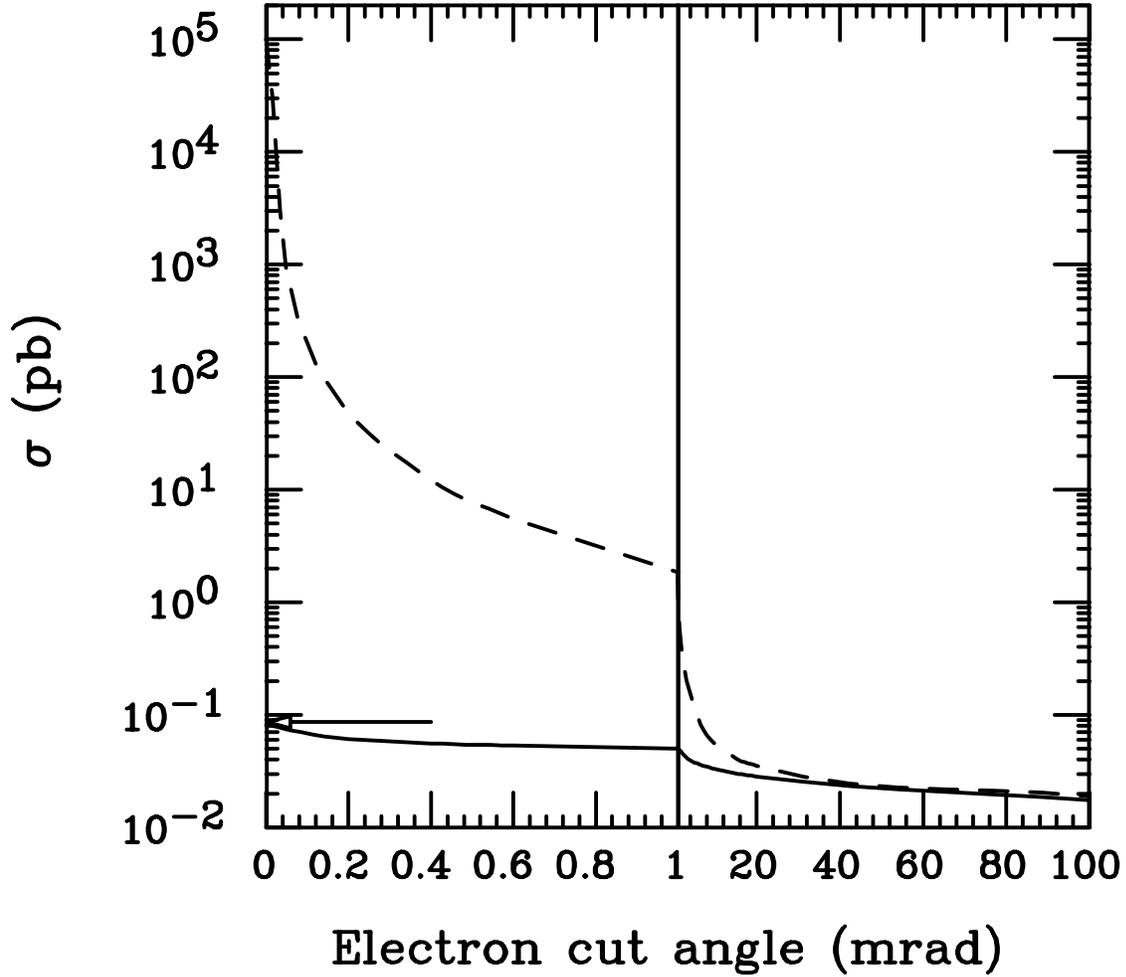}}
\vskip 0.5cm
\caption{The cross-section vs. electron cut angle
for the sub-set of $\gamma$-$W$ diagrams only
at $\surd s =180$ GeV.
The left half of the figure is a magnified view of the small
angle region.
The dashed line shows a result
with a naive Breit-Wigner form for the $W$-propagator and
the solid line corresponds to the introduction of the width in a
gauge-invariant way using the so-called {\it preserved gauge scheme}
as explained in the text.
The arrow shows the cross-section using
the second method described in the section 3.
}
\end{center}
\end{figure}
%%%%%%%%%%%%%%%%%%%%%%%% fig 4 %%%%%%%%%%%%%%%%%%%%%%%%%%%%%%%%%%%%%%%%
\begin{figure}
\begin{center}
\mbox{\epsfig{file=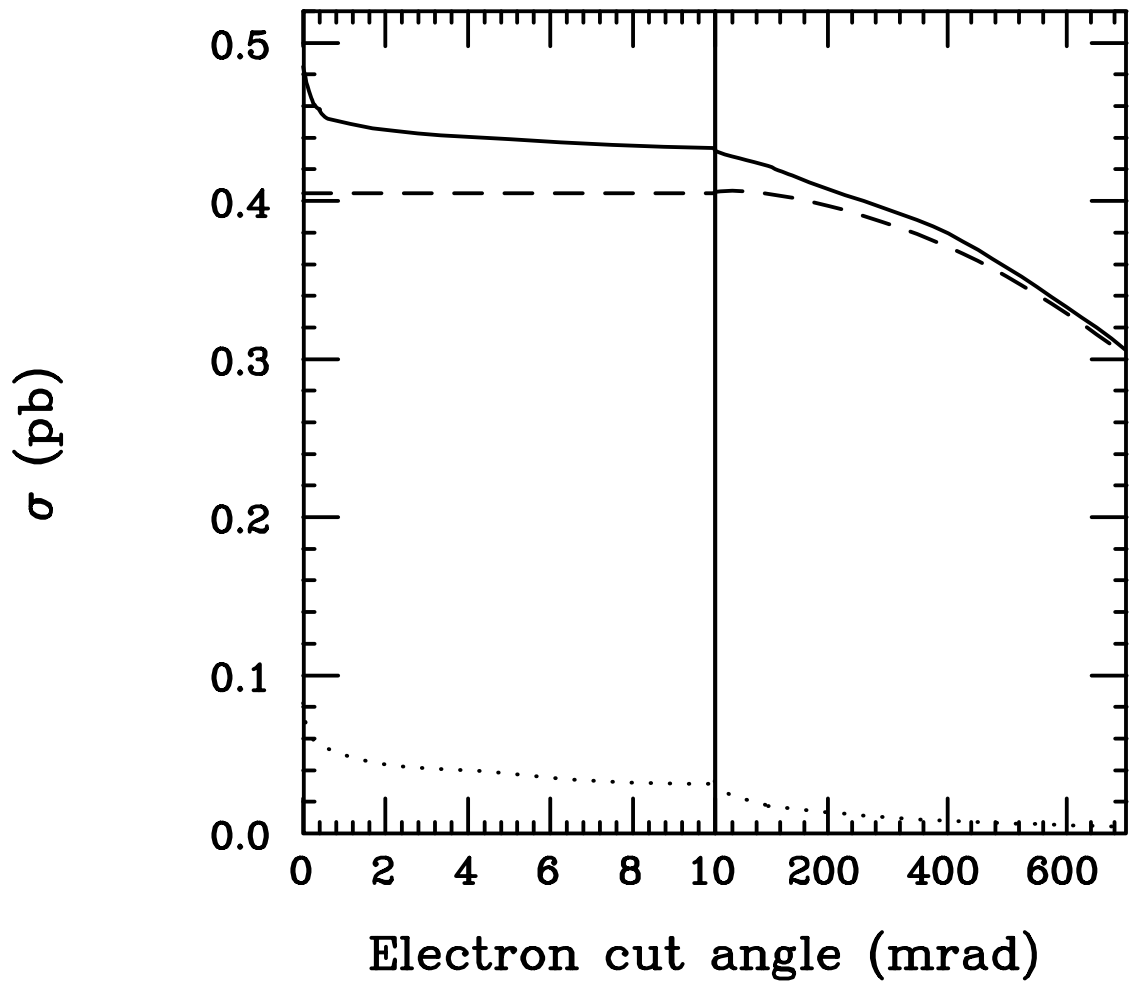}}
\vskip 0.5cm
\caption{The cross-section vs. the $e^-$ cut angle at $\surd s =180$ GeV.
The solid line represents the contribution from all diagrams,
the dashed line from
the double-resonant diagrams, and the dotted line from the $\gamma$-$W$
diagrams.
}
\end{center}
\end{figure}
%%%%%%%%%%%%%%%%%%%%%%%% fig 5 %%%%%%%%%%%%%%%%%%%%%%%%%%%%%%%%%%%%%%%%
\begin{figure}
\begin{center}
\mbox{\epsfig{file=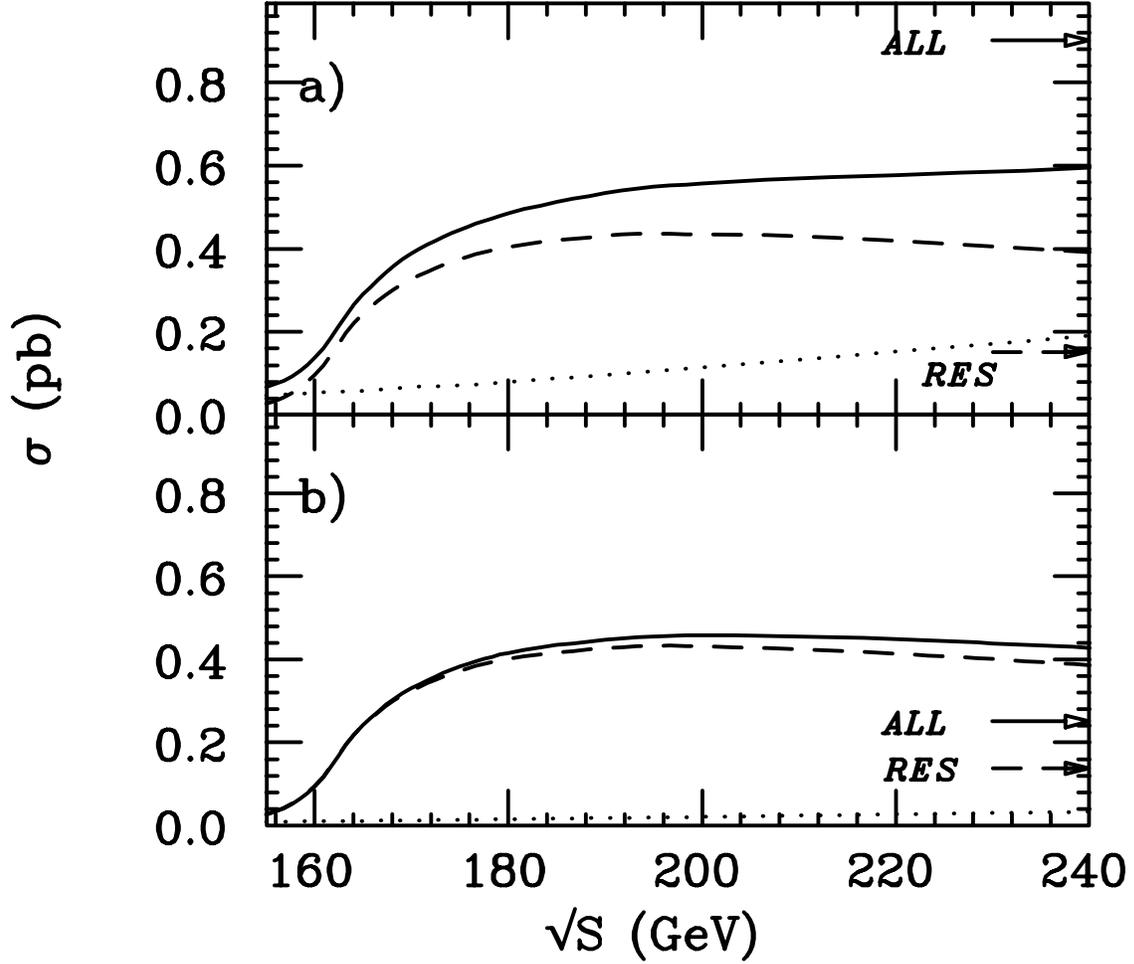}}
\vskip 0.5cm
\caption{The threshold behavior of the total cross-section of the
$e^+e^- \to e^- \bar\nu_e u \bar{d}$ process. Cuts described
in section 2 are applied on $u$ and ${\bar d}$ quarks, a)
no cut on the final electron (case-a),
and b) $\theta_e > 8^\circ$ (case-b).
The solid line shows the cross-section from all diagrams,
the dashed line from the double-resonant diagrams,
and the dotted line from the $t$-channel diagrams.
Results at $\surd s =500$ GeV are shown by solid arrows
(all diagrams) and dashed arrows (double-resonant diagrams).
}
\end{center}
\end{figure}
%%%%%%%%%%%%%%%%%%%%%%%% fig 6 %%%%%%%%%%%%%%%%%%%%%%%%%%%%%%%%%%%%%%%%
\begin{figure}
\begin{center}
\mbox{\epsfig{file=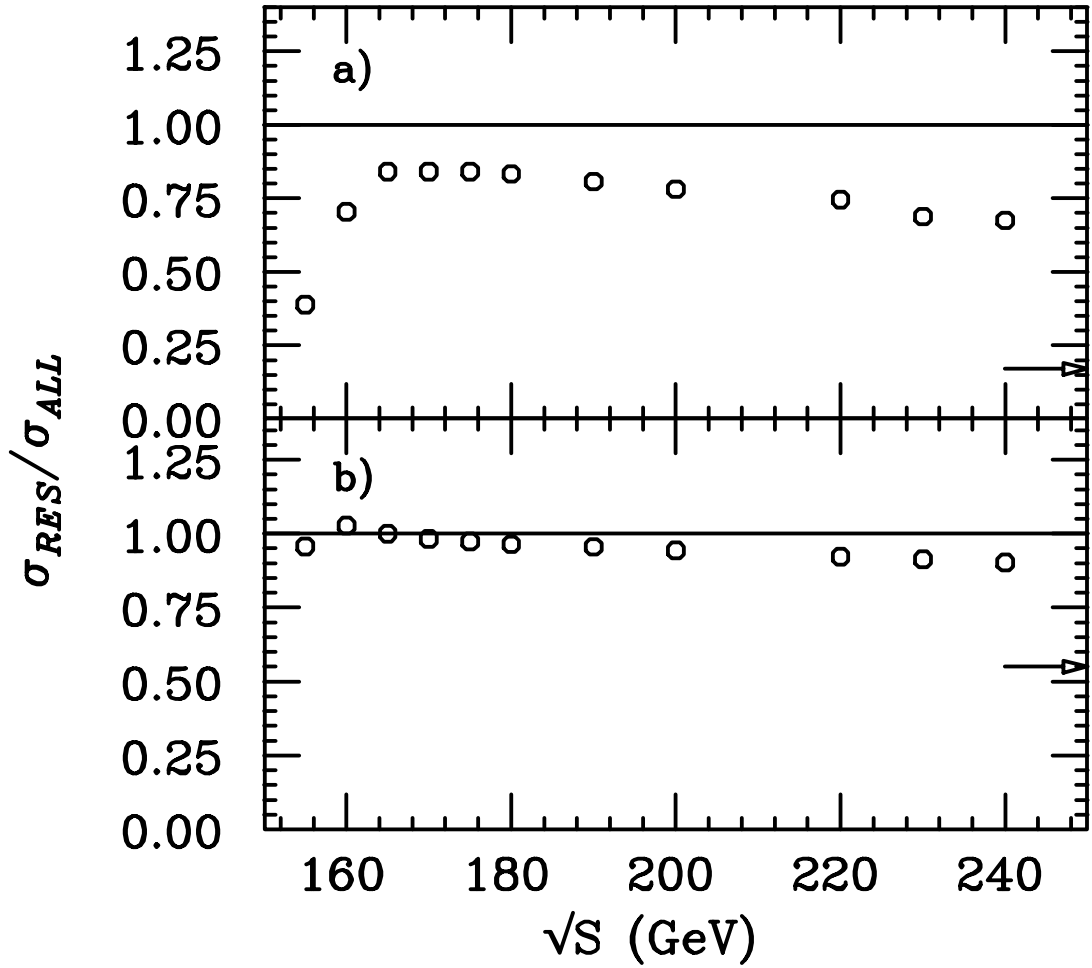}}
\vskip 0.5cm
\caption{The relative contribution of the
double-resonant diagrams to the cross-sections at $\surd s =180$ GeV.
In a), no cut on the out-going electron but only on $u$ and $\bar d$
(case-a) and in b)
an additional cut, $\theta_e > 8^\circ$ (case-b), is imposed.
Results at $\surd s =500$ GeV are shown by arrows.
}
\end{center}
\end{figure}
%%%%%%%%%%%%%%%%%%%%%%%% fig 7 %%%%%%%%%%%%%%%%%%%%%%%%%%%%%%%%%%%%%%%%
\begin{figure}
\begin{center}
\mbox{\epsfig{file=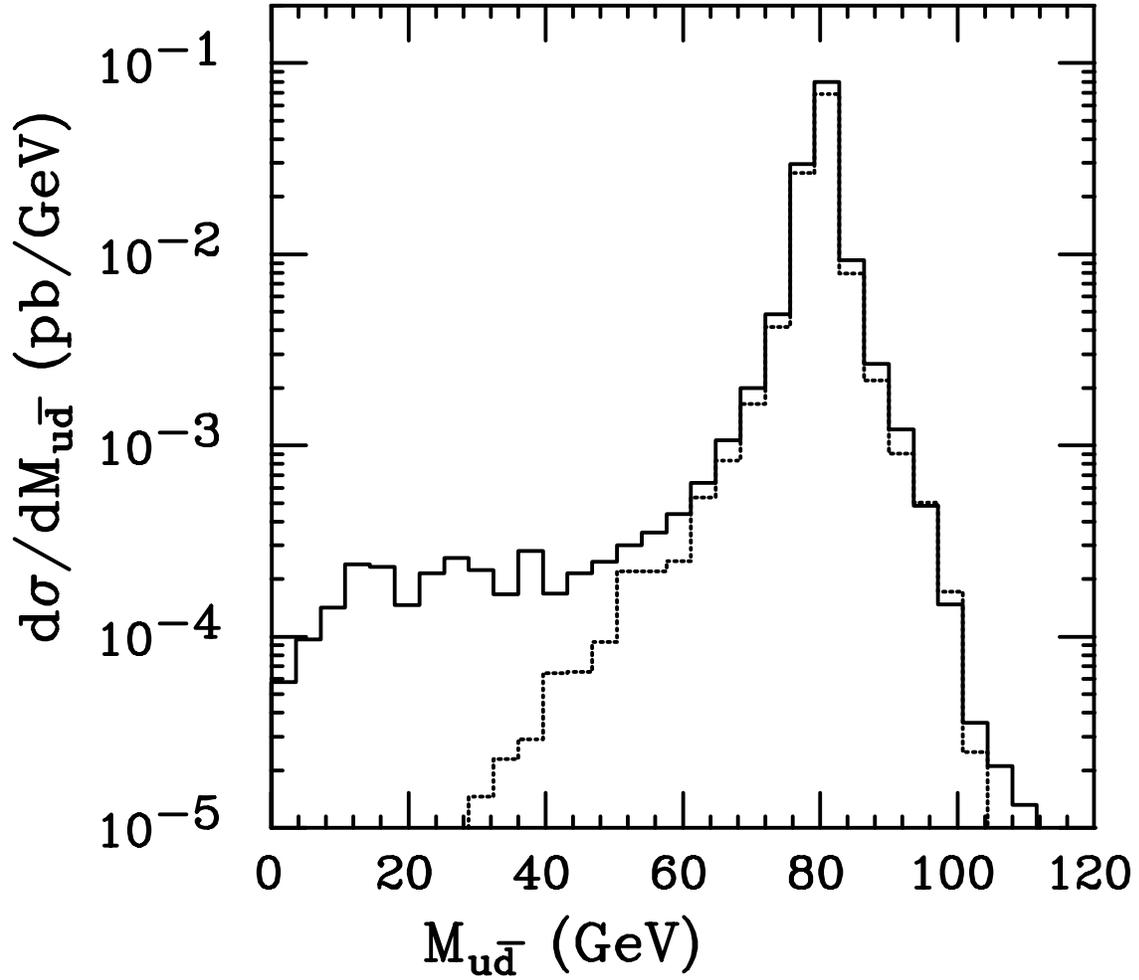}}
\vskip 0.5cm
\caption{The invariant mass distribution of $u$ and $\bar d$ calculated
at $\surd s =180$ GeV. The solid line
(case-a),
and the dotted one with
additional cut, $\theta_e > 8^\circ$(case-b).
}
\end{center}
\end{figure}
%%%%%%%%%%%%%%%%%%%%%%%% fig 8 %%%%%%%%%%%%%%%%%%%%%%%%%%%%%%%%%%%%%%%%
\begin{figure}
\begin{center}
\mbox{\epsfig{file=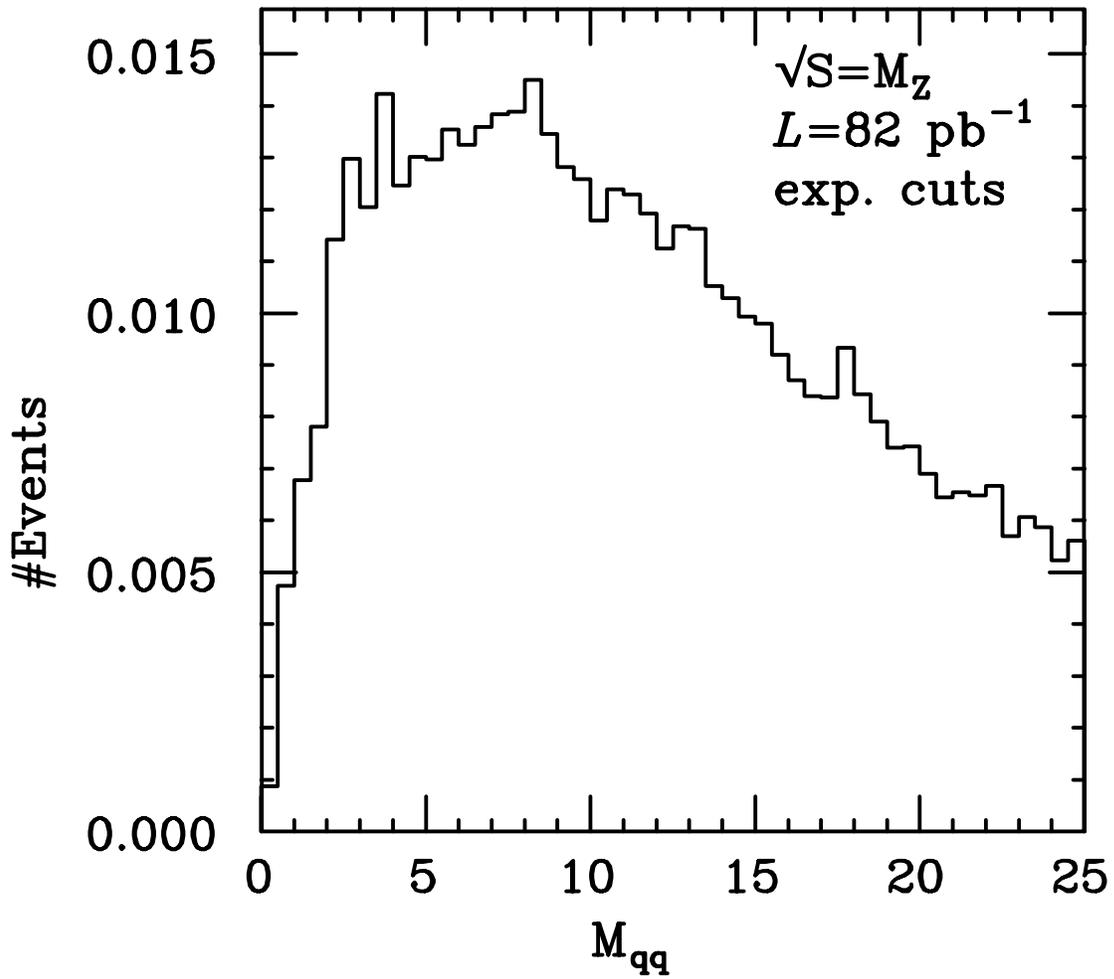}}
\vskip 0.5cm
\caption{The invariant mass distribution of $q$ and ${\bar q}'$ calculated
at $\surd s =M_Z$. The integrated luminosity and experimental cuts
described in ref.[11] but applied at the generator level is assumed.
}
\end{center}
\end{figure}
%%%%%%%%%%%%%%%%%%%%%%%% fig 9 %%%%%%%%%%%%%%%%%%%%%%%%%%%%%%%%%%%%%%%%
\begin{figure}
\begin{center}
\mbox{\epsfig{file=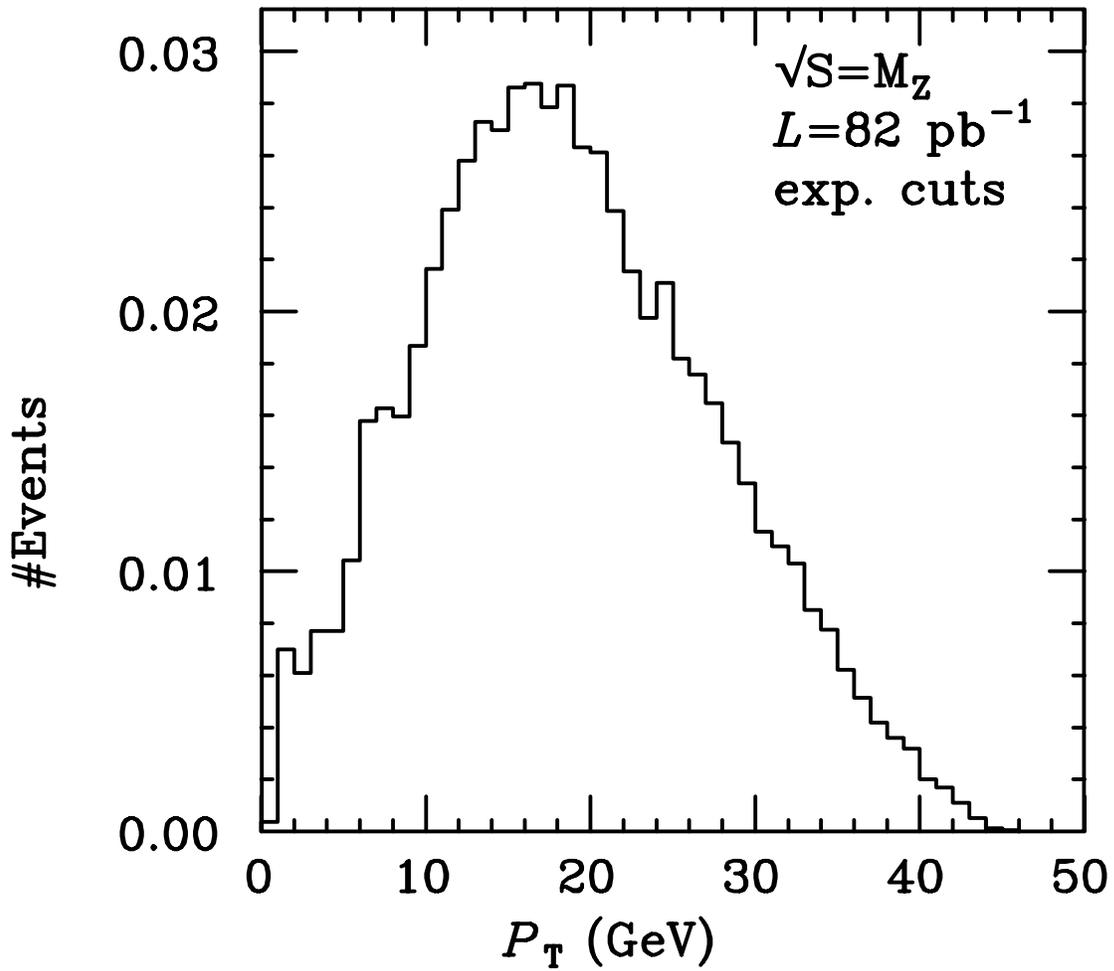}}
\vskip 0.5cm
\caption{The transverse momentum of $q {\bar q}'$ system calculated
at $\surd s =M_Z$. The integrated luminosity and experimental cuts
described in ref.[11] but applied at the generator level are assumed.
}
\end{center}
\end{figure}
\end{document}